\documentclass[%
  reprint,
  showpacs,preprintnumbers,
  amsmath,amssymb,
  aps,
  prb,
]{revtex4-1}

\usepackage{graphicx}
\usepackage{dcolumn}
\usepackage{bm}
\usepackage{hyperref}
\usepackage{multirow}
\usepackage{color}
\usepackage{ulem}
\usepackage{times}
\usepackage{booktabs}

\begin{document}

\preprint{APS/123-QED}

\title{
Effect of magnetic anisotropy on Skyrmions with a high topological number \\ in itinerant magnets
}

\author{Satoru Hayami$^{1}$ and Yukitoshi Motome$^2$}
\affiliation{
$^1$Faculty of Science, Hokkaido University, Sapporo 060-0810, Japan\\
$^2$Department of Applied Physics, University of Tokyo, Tokyo 113-8656, Japan
}
 
\begin{abstract}
We report our numerical results for the effect of magnetic anisotropy on a Skyrmion crystal with a high topological number of two, which was recently discovered in an itinerant electron model [R. Ozawa, S. Hayami, and Y. Motome, Phys. Rev. Lett. \textbf{118}, 147205 (2017)]. 
By performing numerical simulations based on the kernel polynomial method and the Langevin dynamics for the Kondo lattice model on a triangular lattice, we find that the topological property remains robust against the single-ion anisotropy, while the magnetic texture is deformed continuously. 
The resultant spin structure is characterized by three wave numbers (triple-$Q$ state), in which the $xy$ component of spins forms a magnetic vortex crystal and the $z$ component of spins behaves an sinusoidal wave. 
For larger anisotropy, we show that the system exhibits a phase transition from the Skyrmion crystal to topologically trivial phases with vanishing scalar chirality: a single-$Q$ collinear and double-$Q$ noncoplanar states for the easy-axis and easy-plane anisotropy, respectively. 
We also examine the effect of the single-ion anisotropy in an external magnetic field, and find that the field range of the Skyrmion crystal is rather insensitive to the anisotropy, in contrast to another Skyrmion crystal with the topological number of one whose field range is considerably extended (reduced) by the easy-axis (easy-plane) anisotropy. 

\end{abstract}
\maketitle

\section{Introduction}
\label{sec:Introduction}

Spin scalar chirality, which is defined by a triple scalar product of three spins as $\mathbf{S}_i \cdot (\mathbf{S}_j \times \mathbf{S}_k) $, has attracted much interest in condensed matter physics. 
It generates an emergent electromagnetic field for electrons 
through the spin Berry phase mechanism, which has great potential for control of electronic states and transport phenomena, such as the anomalous Hall effect called the topological Hall effect~\cite{Loss_PhysRevB.45.13544,Ye_PhysRevLett.83.3737}. 
In particular, when the scalar chirality acquires a net component by a periodic noncoplanar spin texture, the coherent Berry phase can lead to quantized topological Hall effects~\cite{Ohgushi_PhysRevB.62.R6065,Martin_PhysRevLett.101.156402,Akagi_JPSJ.79.083711,Shindou_PhysRevLett.87.116801}.  

Skyrmion crystals (SkXs) are one of the most attracting examples to exhibit the topological Hall effect~\cite{Bogdanov89,Bogdanov94,Muhlbauer_2009Skyrmion,yu2010real,Yu2011,nagaosa2013topological}. 
They are given by periodic arrangement of swirling spin textures called Skyrmions, and are often found in magnets with chiral lattice structures. 
Such SkXs are stabilized in a magnetic field under competition between the ferromagnetic exchange and the Dzyaloshinskii-Moriya interactions~\cite{dzyaloshinsky1958thermodynamic,moriya1960anisotropic}, the latter of which originates from the relativistic spin-orbit coupling under inversion symmetry breaking~\cite{bak1980theory,rossler2006spontaneous,Yi_PhysRevB.80.054416}. 
Another mechanism for stabilizing SkXs has been investigated in frustrated magnets with competing nearest-neighbor ferromagnetic and further-neighbor antiferromagnetic exchange interactions~\cite{Okubo_PhysRevLett.108.017206,leonov2015multiply,Lin_PhysRevB.93.064430,Hayami_PhysRevB.93.184413,Hayami_PhysRevB.94.174420,Sutcliffe_PhysRevLett.118.247203}. 
In these cases, the SkXs show a common feature for magnetic anisotropy: 
they become more robust by introducing easy-axis anisotropy~\cite{Butenko_PhysRevB.82.052403,Wilson_PhysRevB.89.094411,leonov2015multiply,Lin_PhysRevB.93.064430,Hayami_PhysRevB.93.184413,Batista2016}, while they are destabilized by easy-plane anisotropy~\cite{Lin_PhysRevB.91.224407,leonov2015multiply}.  

Yet another mechanism for SkXs has been developed mainly from the theoretical side, by fully taking into account itinerant nature of electrons. 
The argument is based on the Kondo-type exchange coupling between itinerant electrons and localized spins, which gives rise to effective multiple-spin interactions between localized spins~\cite{Akagi_PhysRevLett.108.096401,Hayami_PhysRevB.90.060402,Ozawa_doi:10.7566/JPSJ.85.103703,Hayami_PhysRevB.95.224424,Hayami_PhysRevB.94.024424} in addition to the conventional Ruderman-Kittel-Kasuya-Yosida (RKKY) interaction~\cite{Ruderman,Kasuya,Yosida1957}. 
Such effective interactions result in noncoplanar spin textures on a variety of lattice geometries: triangular~\cite{Martin_PhysRevLett.101.156402,Akagi_JPSJ.79.083711,Kumar_PhysRevLett.105.216405,Kato_PhysRevLett.105.266405,hayami_PhysRevB.91.075104,Venderbos_PhysRevLett.108.126405,Venderbos_PhysRevB.93.115108,sorn2018field}, 
honeycomb~\cite{Jiang_PhysRevLett.114.216402,Venderbos_PhysRevB.93.115108}, kagome~\cite{Barros_PhysRevB.90.245119,Ghosh_PhysRevB.93.024401}, square~\cite{hayami2018multiple,Hayami_PhysRevLett.121.137202,okada2018multiple}, cubic~\cite{Hayami_PhysRevB.89.085124}, face-centered-cubic~\cite{Shindou_PhysRevLett.87.116801}, pyrochlore~\cite{Chern_PhysRevLett.105.226403}, and Shastry-Sutherland lattices~\cite{Shahzad_PhysRevB.96.224402}. 
This mechanism has two interesting features: 
(i) it does not necessarily require either inversion symmetry breaking or the spin-orbit coupling, and 
(ii) it can produce unconventional SkXs that have never been seen in other mechanisms. 
For the latter, for instance, a recent theoretical study on the triangular lattice revealed a SkX with the topological number of two  ($n_{\rm sk}=2$) at zero magnetic field, and phase transitions with successive changes of the topological number $n_{\rm sk}=2 \to 1 \to 0$ while increasing an external magnetic field~\cite{Ozawa_PhysRevLett.118.147205}. 
However, such a SkX with $n_{\rm sk}=2$ has not been identified in experiments yet. 
Toward experimental observations, it is desirable to examine how it responds to perturbations, such as magnetic anisotropy.  
It will also be helpful for clarifying the similarity and difference between conventional Skyrmions and those rooted in itinerant electrons. 

In the present study, we investigate the effect of single-ion anisotropy on the SkXs with $n_{\rm sk}=2$ and $1$ in itinerant magnets. 
We examine how the spin structures are modulated and how these topological phases are robust against introducing easy-axis or easy-plane anisotropy. 
By performing large-scale Langevin dynamics simulations enabled by the kernel polynomial method (KPM-LD)~\cite{Barros_PhysRevB.88.235101} for the Kondo lattice model on a triangular lattice, we find that the spin structure of the $n_{\rm sk}=2$ SkX is deformed into an anisotropic form composed of magnetic vortices in the $xy$ spin component and a sinusoidal wave in the $z$ spin component. 
In addition, we show that the SkX shows a topological trivial-nontrivial transition to a single-$Q$ (1$Q$) collinear 
[double-$Q$ (2$Q$) noncoplanar] state while increasing the easy-axis (easy-plane) anisotropy. 
We also compare the robustness of the SkXs with $n_{\rm sk}=2$ and $1$ in an applied magnetic field. 
We find that the field range of the $n_{\rm sk}=2$ state is rather insensitive to the anisotropy, while the $n_{\rm sk}=1$ state is substantially stabilized (destabilized) by the easy-axis (easy-plane) anisotropy similar to conventional SkXs found in other systems. 

The rest of the paper is organized as follows. 
In Sec.~\ref{sec:Model and Method}, we introduce the Kondo lattice model including the single-ion anisotropy and the Zeeman coupling to an external magnetic field, outline the KPM-LD method, and define the observables that we evaluate. 
We examine the effect of the single-ion anisotropy on the $n_{\rm sk}=2$ SkX at zero magnetic field and for a nonzero field in Sec.~\ref{sec:Effect of single-ion anisotropy}. 
Section~\ref{sec:Summary} is devoted to a summary. 

\section{Model and Method}
\label{sec:Model and Method}

\subsection{Kondo lattice model}
\label{sec:Kondo lattice model}
We consider the Kondo lattice model including the effect of the single-ion anisotropy and the external magnetic field on the triangular lattice. 
The Hamiltonian is given by  
\begin{align}
\label{eq:Ham}
\mathcal{H} = 
-\sum_{i, j,  \sigma} t_{ij} c^{\dagger}_{i\sigma}c_{j \sigma}
+J \sum_{i} \mathbf{s}_i \cdot \mathbf{S}_i - A \sum_i (S_i^z)^2 - H \sum_i S_i^z. 
\end{align}
The first term represents the kinetic energy of itinerant electrons, where $c^{\dagger}_{i\sigma}$ ($c_{i \sigma}$) is a creation (annihilation) operator of an itinerant electron at site $i$ and spin $\sigma$. 
The second term in Eq.~(\ref{eq:Ham}) represents the onsite exchange coupling between itinerant electron spins $\mathbf{s}_i=(1/2)\sum_{\sigma, \sigma'}c^{\dagger}_{i\sigma} \bm{\sigma}_{\sigma \sigma'} c_{i \sigma'}$ and localized spins $\mathbf{S}_i$ with coupling constant $J$, where $\bm{\sigma}=(\sigma^x,\sigma^y,\sigma^z)$ is the vector of Pauli matrices. 
We regard $\mathbf{S}_i$ as a classical spin with fixed length $|\mathbf{S}_i|=1$ (the sign of $J$ is irrelevant).
The third and fourth terms describe the easy-axis ($A>0$) or easy-plane ($A<0$) anisotropy and the Zeeman coupling to an external magnetic field along the $z$ direction, respectively, both of which are taken into account only for the localized spins for simplicity.

The ground state of the model in Eq.~(\ref{eq:Ham}) was investigated in the absence of the single-ion anisotropy ($A=0$)~\cite{Ozawa_PhysRevLett.118.147205}. 
The system exhibits a triple-$Q$ (3$Q$) SkX with $n_{\rm sk}=2$ at zero field, which is characterized by three wave numbers $\mathbf{Q}_\eta=R(2\pi(\eta-1)/3)(\pi/3,0)$ ($\eta=1$-$3$), where $R(\theta)$ represents the rotational operation around the $z$ axis by $\theta$. 
Furthermore, while increasing the magnetic field, topological phase transitions occur successively as $n_{\rm sk }=2 \to 1 \to 0$. 
In the following calculations, we examine the effect of the single-ion anisotropy on the SkXs with $n_{\rm sk}=2$ and $1$ by introducing the third term in Eq.~(\ref{eq:Ham}) with setting the other parameters at the same values as those in Ref.~\onlinecite{Ozawa_PhysRevLett.118.147205}: 
the nearest-neighbor and third-neighbor hoppings, $t_1=1$ and $t_3=-0.85$, respectively, $J=1$, and the chemical potential $\mu=-3.5$. 

\subsection{Simulation method}
\label{sec:Langevin dynamics simulation}

We investigate the ground state of the Kondo lattice model in Eq.~(\ref{eq:Ham}) by performing the KPM-LD simulation, which is an unbiased numerical simulation based on Langevin dynamics combined with the kernel polynomial method~\cite{Weis_RevModPhys.78.275,Barros_PhysRevB.88.235101}.  
This method enables the calculations for large system sizes, typically up to $10^4$ sites, and has been applied to similar models with itinerant electrons~\cite{Barros_PhysRevB.88.235101,Barros_PhysRevB.90.245119,Ozawa_doi:10.7566/JPSJ.85.103703,Ozawa_PhysRevLett.118.147205,Hayami_PhysRevB.94.024424,Wang_PhysRevLett.117.206601,Ozawa_PhysRevB.96.094417,Chern_PhysRevB.97.035120}.   
Our simulation is done at zero temperature from initial states with random spin configurations for a $96^2$-site cluster of the triangular lattice with periodic boundary conditions in both directions. 
In the kernel polynomial method, we expand the density of states by up to 2000th order of Chebyshev polynomials with $16^2$ random vectors~\cite{Tang12}. 
In the Langevin dynamics, we use a projected Heun scheme~\cite{Mentink10} for 1000-5000 steps with the time interval  $\Delta \tau =2$.  

\subsection{Physical observables}
\label{sec:Physical observables}

For spin configurations obtained by the KPM-LD simulation, we calculate the spin structure factor to identify each magnetic phase, which is given by 
\begin{align}
S(\mathbf{q})=S^{xx} (\mathbf{q})+ S^{yy} (\mathbf{q})+S^{zz} (\mathbf{q}), 
\end{align}
where 
\begin{align}
S^{\alpha\alpha} (\mathbf{q}) = \frac{1}{N} \sum_{j,l} S_j^{\alpha}S_l^{\alpha} e^{i \mathbf{q}\cdot (\mathbf{r}_j-\mathbf{r}_l)}, 
\end{align}
with $\alpha=x,y,z$ and $N$ is the system size. 
We also compute 
\begin{align}
S^{\perp}(\mathbf{q}) = S^{xx} (\mathbf{q})+ S^{yy} (\mathbf{q}). 
\end{align}
In addition, we introduce the following notation for the magnetic moments with wave number $\mathbf{q}$, 
\begin{align}
m_{\mathbf{q}}=\sqrt{\frac{S(\mathbf{q})}{N}}. 
\label{eq:m_q}
\end{align}
Note that the uniform magnetization is given by $M=m_{\mathbf{q}=\mathbf{0}}$. 
In addition, in order to identify whether the obtained phase is chiral or not, we calculate the net spin scalar chirality of the localized spins, which is defined as 
\begin{align}
\chi_{\rm sc} =  \frac{1}{N} \sum_{p} \chi_{p}=\frac{1}{N} \sum_{p}\mathbf{S}_{i} \cdot (\mathbf{S}_{j} \times \mathbf{S}_{k}), 
\label{eq:chi_sc}
\end{align}
where $\chi_p$ is a local scalar chirality defined on triangle plaquette $p$; 
$i,j,k$ are sites on each triangle plaquette $p$ in the counterclockwise direction.

\section{Results}
\label{sec:Effect of single-ion anisotropy}

In this section, we present numerical results for the effect of the single-ion anisotropy obtained by the KPM-LD method. 
In Sec.~\ref{sec:Zero field}, we show how the SkX with $n_{\rm sk}=2$ is affected by the anisotropy at zero field. 
We extend the study to nonzero fields and compare the robustness of SkXs with $n_{\rm sk}=2$ and $n_{\rm sk}=1$ in Sec.~\ref{sec:Magnetic field}. 

\subsection{At zero field}
\label{sec:Zero field}

\begin{figure}[htb!]
\begin{center}
\includegraphics[width=1.0 \hsize]{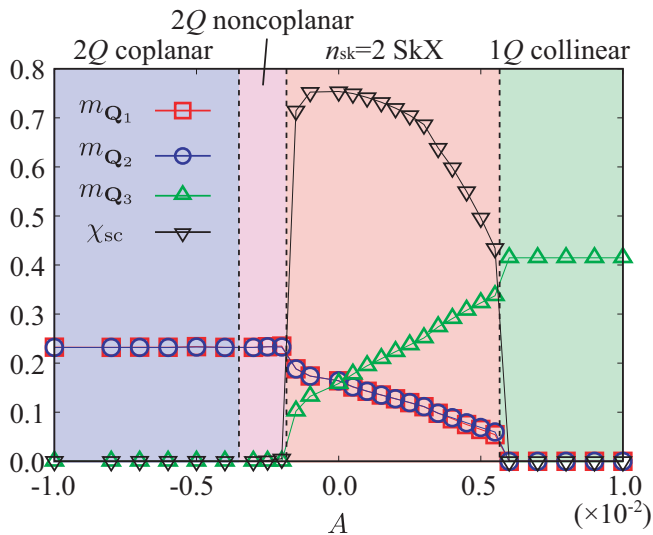} 
\caption{
\label{Fig:sq_Adep}
KPM-LD results for the model in Eq.~(\ref{eq:Ham}) at zero field: the single-ion anisotropy $A$ dependences of the $\mathbf{Q}_\nu$ components of the magnetization [Eq.~(\ref{eq:m_q})] and the net spin scalar chirality [Eq.~(\ref{eq:chi_sc})]. 
The vertical dashed lines show the phase boundaries. 
}
\end{center}
\end{figure}

\begin{figure*}[htb!]
\begin{center}
\includegraphics[width=0.95 \hsize]{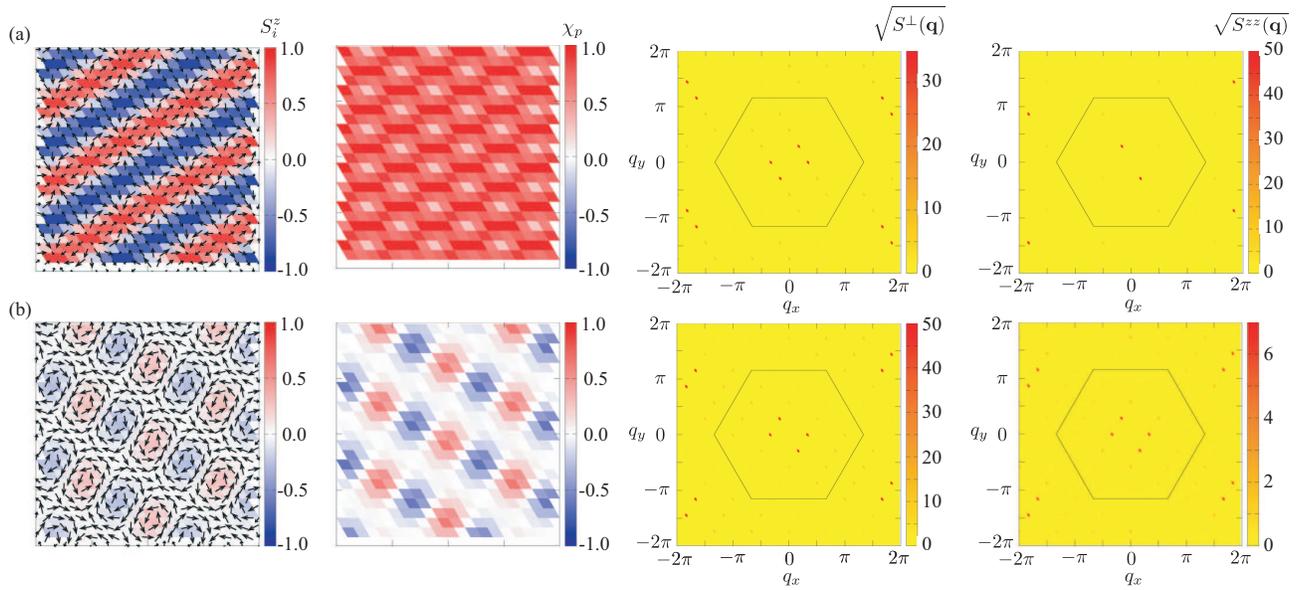} 
\caption{
\label{Fig:spinpattern_chiral}
(Leftmost) Snapshots of the spin configurations in (a) the 3$Q$ SkX with $n_{\rm sk}=2$ for $A=2.5 \times 10^{-3}$ and (b) the 2$Q$ noncoplanar state for $A=-2.5 \times 10^{-3}$ at $H=0$.  
The contour shows the $z$ component of the spin moment. 
(Middle left) 
Snapshots of the spin scalar chirality. 
(Middle right and rightmost) 
The square root of the $xy$ and $z$ components of the spin structure factor, respectively. 
In the right two columns, the hexagons represent the first Brillouin zone. 
}
\end{center}
\end{figure*}

\begin{figure}[htb!]
\begin{center}
\includegraphics[width=1.0 \hsize]{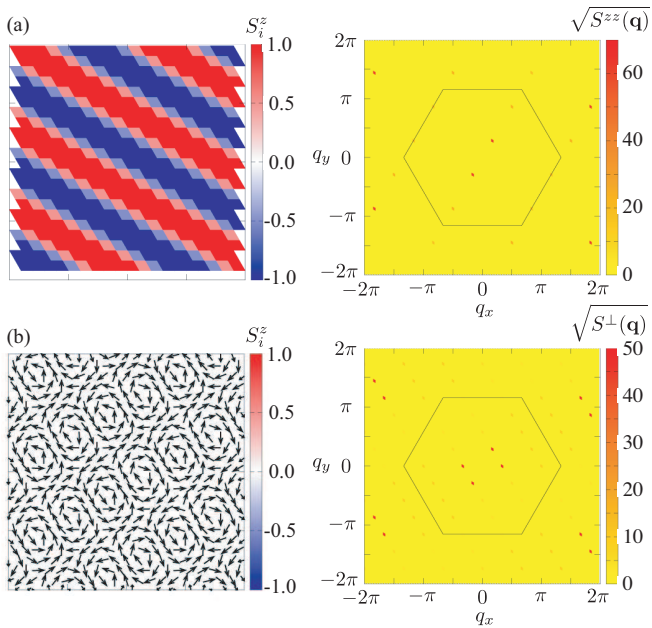} 
\caption{
\label{Fig:spinpattern_nochiral}
(Left) Snapshots of the spin configurations in (a) the 1$Q$ collinear state for $A=8 \times 10^{-3}$ and (b) the 2$Q$ coplanar state for $A=-6 \times 10^{-3}$ at $H=0$. 
The contour shows the $z$ component of the spin moment. 
(Right) 
The square root of the $z$ ($xy$) component of the spin structure factor in the upper panel (lower). 
The $xy$ ($z$) component of the spin structure factor is negligible in (a) [(b)]. 
In the right column, the hexagons represent the first Brillouin zone. 
}
\end{center}
\end{figure}

First, we present the KPM-LD results at zero field $H=0$. 
Figure~\ref{Fig:sq_Adep} shows the $\mathbf{Q}_\nu$ component of the magnetization, $m_{\mathbf{Q}_\nu}$ [Eq.~(\ref{eq:m_q})], and the net spin scalar chirality $\chi_{\rm sc}$ [Eq.~(\ref{eq:chi_sc})], as functions of the single-ion anisotropy $A$. 
At $A=0$, the 3$Q$ SkX with $n_{\rm sk}=2$ is realized as discussed in Ref.~\onlinecite{Ozawa_PhysRevLett.118.147205}. 
In this state, the amplitudes of the $\mathbf{Q}_1$, $\mathbf{Q}_2$, and $\mathbf{Q}_3$ components are equivalent, i.e., $m_{\mathbf{Q}_1} = m_{\mathbf{Q}_2} = m_{\mathbf{Q}_3}$, as shown in Fig.~\ref{Fig:sq_Adep}. 
At the same time, the noncoplanar spin structure leads to the nonzero value of $\chi_{\rm sc}$; the resultant topological number is quantized at two~\cite{Ozawa_PhysRevLett.118.147205}. 
Note that the helicity of the $n_{\rm sk}=2$ SkX at $A=H=0$ is arbitrary because of spin rotational symmetry. 

When the single-ion anisotropy is introduced ($A\neq 0$), the $xy$ and $z$ components of the magnetization behave differently due to the breaking of rotational symmetry in spin space. 
The $xy$ component shows the 2$Q$ structures with equal intensities at $\mathbf{Q}_1$ and $\mathbf{Q}_2$, while the $z$ component shows the 1$Q$ structure at $\mathbf{Q}_3$, as shown in Fig.~\ref{Fig:spinpattern_nochiral}. 
We show the snapshots of the spin configuration in Fig.~\ref{Fig:spinpattern_chiral}(a). 
(The choice of $\mathbf{Q}_\nu$ for 2$Q$ and 1$Q$ depends on the snapshot.) 
The intensities of the 2$Q$ peaks in the $xy$ component are smaller (larger) than that of the 1$Q$ peak in the $z$ component for the easy-axis anisotropy $A>0$ (easy-plane anisotropy $A<0$). 
We note that the topological number remains unchanged at $n_{\rm sk}=2$ for the continuous modulations of the spin structures by $A$. 
From these observations, we find that the real-space spin configuration of the 3$Q$ SkX with $n_{\rm sk}=2$ is approximately given by  
\begin{align}
\label{eq:tripleQ}
\mathbf{S}_i \propto [\cos (\mathbf{Q}_1\cdot \mathbf{r}_i),\cos (\mathbf{Q}_2\cdot \mathbf{r}_i),a^z \cos (\mathbf{Q}_3\cdot \mathbf{r}_i)],  
\end{align}
where the coefficient $a^z$ depends on $A$: $a^z>1$ ($a^z<1$) for $A>0$ ($A<0$). 
Note that the $xy$ component in Eq.~(\ref{eq:tripleQ}) is generally described by an arbitrary linear combination of $S^x_i$ and $S^y_i$ owing to the rotational symmetry around the $z$ axis. 
Thus, the spin texture is modified by the single-ion anisotropy into a superposition of the 2$Q$ vortices in the $xy$ component and a sinusoidal wave in the $z$ component. 
The helicity for the 2$Q$ vortices is arbitrary, which reflects the presence of the rotational symmetry in the $xy$ component of spins. 
Interestingly, the real-space spin structures lack clear Skyrmion cores, in stark contrast to conventional SkXs, as shown in Fig.~\ref{Fig:spinpattern_chiral}(a).

While increasing the easy-axis anisotropy, the $z$ component is developed almost linearly to $A$, while the $x$ and $y$ components are suppressed, as shown in Fig.~\ref{Fig:sq_Adep}.  
Simultaneously, the net scalar chirality decreases monotonically by increasing $A$, which indicates that the solid angle spanned by three spins becomes smaller for larger $A$. 
At $A\sim 0.006$, the SkX with $n_{\rm sk}=2$ turns into a 1$Q$ collinear state. 
This is a first-order transition with vanishing spin scalar chirality, as shown in Fig.~\ref{Fig:sq_Adep}. 
In this 1$Q$ collinear phase, the $z$ component of the spin structure factor shows a 1$Q$ peak, while the $xy$ component is negligibly small, as shown in Fig.~\ref{Fig:spinpattern_nochiral}(a). 

Meanwhile, for the easy-plane anisotropy $A<0$, the $n_{\rm sk}=2$ SkX turns into a 2$Q$ noncoplanar state at $A\sim -0.002$, as shown in Fig.~\ref{Fig:sq_Adep}. 
The typical spin configuration is shown in Fig.~\ref{Fig:spinpattern_chiral}(b). 
In this state, the spin configuration is characterized by the 2$Q$ modulation: 
both the $xy$ and $z$ components of the spin structure factor have two dominant peaks, while the intensities of the $xy$ component is much larger than that of the $z$ component in Fig.~\ref{Fig:spinpattern_chiral}(b). 
The net scalar chirality is zero in this phase, although there is a 2$Q$ chiral density wave [see the middle left panel of Fig.~\ref{Fig:spinpattern_chiral}(b)]. 
While further increasing the easy-plane anisotropy, the small $z$ component of the magnetization is suppressed to zero and the system turns into a 2$Q$ coplanar state at $A\sim-0.0035$, whose spin structure is shown in Fig.~\ref{Fig:spinpattern_nochiral}(b).  

\subsection{In a magnetic field}
\label{sec:Magnetic field}

\begin{figure}[htb!]
\begin{center}
\includegraphics[width=0.85 \hsize]{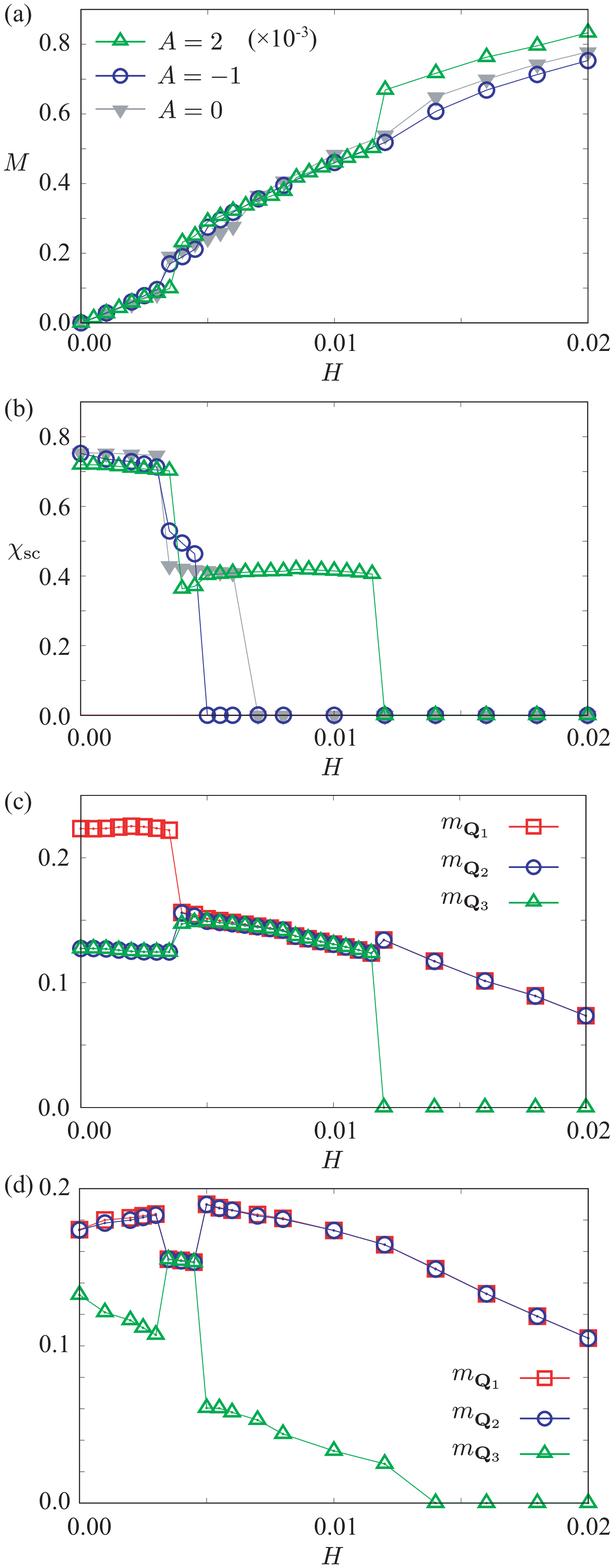} 
\caption{
\label{Fig:magcurve_z}
(a) Magnetization curves and (b) the net scalar chirality for $A=0.002$ and $-0.001$ under the magnetic field $H$. 
The data at $A=0$ are also shown for comparison. 
(c), (d) $H$ dependences of $m_{\mathbf{Q}_\nu}$ ($\nu=1-3$) for (c) $A=0.002$ and (d) $A=-0.001$. 
}
\end{center}
\end{figure}

\begin{figure*}[htb!]
\begin{center}
\includegraphics[width=1.0 \hsize]{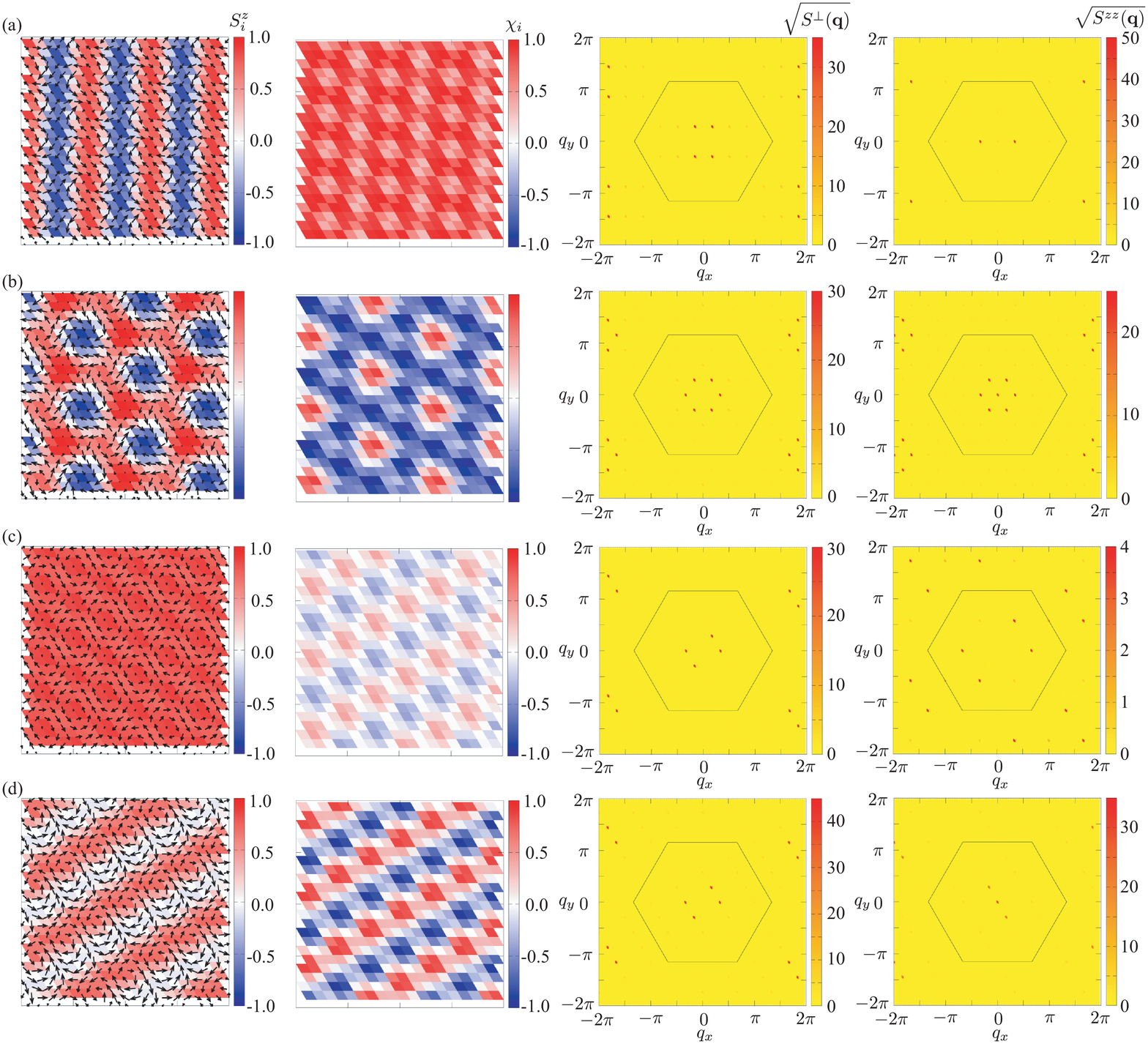} 
\caption{
\label{Fig:spinpattern_magfield_z}
(Leftmost) Snapshots of the spin configurations in (a) the 3$Q$ SkX with $n_{\rm sk}=2$ for $H=0.002$ and $A=0.002$, (b) the 3$Q$ SkX with $n_{\rm sk}=1$ for $H=0.004$ and $A=0.002$, (c) the 2$Q$ state with $n_{\rm sk}=0$ for $H=0.018$ and $A=0.002$, and (d) the 3$Q$ state with $n_{\rm sk}=0$ for $H=0.006$ and $A=-0.001$. 
The contour shows the $z$ component of the spin moment. 
(Middle left) 
Snapshots of the spin scalar chirality. 
(Middle right and rightmost) 
The square root of the $xy$ and $z$ components of the spin structure factor, respectively. 
In the right two columns, the hexagons represent the first Brillouin zone. 
In (c), the field-induced $\mathbf{q}=\mathbf{0}$ component is subtracted for clarity. 
}
\end{center}
\end{figure*}

Next, we discuss the effect of the magnetic field on the $n_{\rm sk}=2$ SkX in the presence of the single-ion anisotropy $A$. 
In the previous study at $A=0$, while increasing the magnetic field $H$, the 3$Q$ SkX with $n_{\rm sk}=2$ changes to another 3$Q$ SkX with $n_{\rm sk}=1$, and finally turns into a topologically-trivial 3$Q$ noncoplanar state with $n_{\rm sk}=0$~\cite{Ozawa_PhysRevLett.118.147205}. 
We here compute the magnetic states at $A\neq 0$ and $H\neq 0$, and examine how the $n_{\rm sk}=2$ SkX is robust against $H$ in comparison with the $n_{\rm sk}=1$ SkX. 
 
Figure~\ref{Fig:magcurve_z}(a) shows the magnetization curves at $A=0.002$ and $A=-0.001$ obtained by the KPM-LD simulations. 
We plot the result at $A=0$ for comparison, which reproduces the previous result~\cite{Ozawa_PhysRevLett.118.147205}. 
In the low field region, the magnetizations continuously increase with $H$, and the spin structures for $A\neq 0$ are similar to that at $A=0$, as exemplified in Fig.~\ref{Fig:spinpattern_magfield_z}(a).  
While further increasing $H$, the magnetizations show a jump at $H\sim 0.003$ almost irrespective of $H$. 
This is the topological phase transition from the 3$Q$ SkX with $n_{\rm sk}=2$ to another 3$Q$ SkX with $n_{\rm sk}=1$. 
In the field-induced SkXs, the cores of swirling spin texture with $S^z_i=-1$ form a triangular lattice, as shown in the snapshot in Fig.~\ref{Fig:spinpattern_magfield_z}(b).  
In this state, the magnetization has equal weights for the $\mathbf{Q}_\nu$ components ($\nu=1,2,3$), as plotted in Figs.~\ref{Fig:magcurve_z}(c) and \ref{Fig:magcurve_z}(d), and correspondingly, the spin structure factors have six peaks with equal intensities, as shown in the right two panels of Fig.~\ref{Fig:spinpattern_magfield_z}(b). 
The $n_{\rm sk}=1$ SkX carries a net scalar chirality, which is reduced to around half from the value in the $n_{\rm sk}=2$ SkX, as plotted in Fig.~\ref{Fig:magcurve_z}(b). 
All these features are similar to the SkX with $n_{\rm sk}=1$ found at $A=0$ in the previous study~\cite{Ozawa_PhysRevLett.118.147205}.
Note that this $n_{\rm sk}=1$ SkX is similar to those reported in frustrated magnets~\cite{Okubo_PhysRevLett.108.017206,leonov2015multiply,Lin_PhysRevB.93.064430,Hayami_PhysRevB.93.184413}, and rather conventional compared to the $n_{\rm sk}=2$ state. 

For a larger magnetic field, the system behaves differently between $A=0.002$ and $A=-0.001$. 
At $A=0.002$, the $n_{\rm sk}=1$ SkX survives up to $H\sim 0.012$ and turns into another state with a jump of the magnetization and vanishment of the spin scalar chirality, as shown in Figs.~\ref{Fig:magcurve_z}(a) and \ref{Fig:magcurve_z}(b). 
The high-field state is a topologically-trivial 2$Q$ noncoplanar state, as indicated in Figs.~\ref{Fig:magcurve_z}(c) and \ref{Fig:spinpattern_magfield_z}(c). 
We note that this state exhibits a 1$Q$ modulation of the spin scalar chirality due to a small peak structure of the $z$-spin component at higher harmonics, as shown in Fig.~\ref{Fig:spinpattern_magfield_z}(c). 
On the other hand, at $A=-0.001$, the phase transition from the $n_{\rm sk}=1$ SkX occurs at a much lower field $H\sim 0.005$ to a topologically-trivial 3$Q$ noncoplanar state, as shown in Figs.~\ref{Fig:magcurve_z}(a), \ref{Fig:magcurve_z}(b), and \ref{Fig:magcurve_z}(d). 
We present the spin structure in Fig.~\ref{Fig:spinpattern_magfield_z}(d); 
it is characterized by a 2$Q$ modulation in the $xy$ component with equal intensities and a 1$Q$ modulation in the $z$ component. 
Meanwhile, the chirality pattern is characterized by the 1$Q$ modulation also in this case. 
With a further increase of $H$, the $z$ component of the magnetization is gradually suppressed and vanishes at $H\sim 0.014$ as shown in Fig.~\ref{Fig:magcurve_z}(d), which indicates the phase transition to the 2$Q$ noncoplanar state. 

Our results indicate that the stability of the SkX states in the presence of the single-ion anisotropy is largely different between the two SkXs. 
Compared to the result at $A=0$, the field region of the $n_{\rm sk} =2$ SkX does not show a notable change for both $A=0.002$ and $A=-0.001$. 
On the other hand, the field range of the $n_{\rm sk}=1$ SkX is substantially extended (reduced) for $A=0.002$ ($A=-0.001$). 
In the case of the easy-plane anisotropy with $A=-0.001$, the $n_{\rm sk}=1$ state is unstable to the intervening 3$Q$ state that does not appear for the easy-axis case with $A=0.002$. 
Similar contrasting response to the easy-axis and easy plane anisotropy is found for SkXs in frustrated magnets~\cite{leonov2015multiply,Lin_PhysRevB.93.064430,Hayami_PhysRevB.93.184413}.

\section{Summary}
\label{sec:Summary}

In summary, we have investigated the stability of the SkXs with different topological numbers ($n_{\rm sk}$) in itinerant magnets against the single-ion anisotropy and external magnetic field by large-scale numerical simulations based on the KPM-LD method for the Kondo lattice model on a triangular lattice. 
We showed that the spin structure of the $n_{\rm sk}=2$ SkX is continuously deformed by the anisotropy into an anisotropic 3$Q$ state composed of magnetic vortices in the $xy$-spin component and a sinusoidal wave in the $z$-spin component. 
Moreover, we found that the system exhibits a topological phase transition from the $n_{\rm sk}=2$ SkX to a 1$Q$ collinear (2$Q$ noncoplanar) state while increasing the easy-axis (easy-plane) anisotropy at zero field.  
We also clarified that the SkXs with $n_{\rm sk}=2$ and 1 show contrasting behaviors in an applied magnetic field in the presence of the single-ion anisotropy; the stable field range of the $n_{\rm sk}=2$ SkX is not much affected by the anisotropy, while the range of the $n_{\rm sk}=1$ SkX is substantially extended (reduced) by the easy-axis (easy-plane) anisotropy. 
Our result underscores that the unconventional SkX with $n_{\rm sk}=2$ is expected to be found in materials with the relatively small spin-charge coupling and small magnetic anisotropy in itinerant magnets. 
Such conditions may be attained in monolayer metals on substrates~\cite{Bergmann_PhysRevLett.96.167203,heinze2011spontaneous,Yoshida_PhysRevLett.108.087205,Zimmermann_PhysRevB.90.115427,Serrate_PhysRevB.93.125424,hoffmann2017antiSkyrmions} and in bulk systems where chiral states were recently reported in SrFeO$_3$~\cite{Ishiwata_PhysRevB.84.054427,ishiwata2018emergent} and Gd$_2$PdSi$_3$~\cite{kurumaji2018skyrmion}. 
More sophisticated analyses, which take account of the multi-orbital degree of freedom and the spin-orbit coupling, are left for future study.

\begin{acknowledgments}
The authors acknowledge R. Ozawa and K. Barros for enlightening discussions in the early stage of this study. 
This research was supported by JSPS KAKENHI Grant No. JP18H04296 (J-Physics) and JP18K13488. Parts of the numerical calculations were performed in the supercomputing systems in ISSP, the University of Tokyo.
\end{acknowledgments}

\bibliographystyle{apsrev}
\bibliography{ref}

\end{document}